\newcommand\cA{{\cal A}}

\newcommand\cC{{\cal C}}

\newcommand\cI{{\cal I}}
\newcommand\cQ{{\cal Q}}

\documentclass[12pt]{article}
\begin{document}

\begin{titlepage}
\nopagebreak
\begin{flushright}
June 2001\hfill
UT-947\\
hep-th/0107007
\end{flushright}

\renewcommand{\thefootnote}{\fnsymbol{footnote}}
\vfill
\begin{center}
{\Large Projection Operators and D-branes}\\
{\Large in Purely Cubic Open String Field Theory}

\vskip 20mm

Yutaka MATSUO\footnote{
{\tt E-mail address: matsuo@phys.s.u-tokyo.ac.jp}
}
\vskip 1cm
Department of Physics, University of Tokyo\\
Hongo 7-3-1, Bunkyo-ku\\
Tokyo 113-0033\\
Japan
\end{center}
\vfill

\begin{abstract}
We study a matrix version of the purely cubic open string
field theory as describing the expansion
around the closed string vacuum.
Any D-branes in the given closed string background can appear
as classical solutions by using the identity projectors.
Expansion around this solution
gives the correct kinetic term
for the open strings on the created D-branes
while there are some subtleties in the unwanted degree of 
freedom.
\end{abstract}
\vfill
\end{titlepage}


\section{Projectors in String Field Theory}
Recently Rastelli, Sen and 
Zwiebach \cite{r-RSZ1, r-RSZ2}
pioneered the open string field theory 
around the nonperturbative (closed string) vacuum.
They conjectured Witten's open string field 
theory action \cite{r-Witten} should be modified
in the following form,
\begin{equation}\label{e-RSZ}
 S=\frac{1}{2}\int \Psi\star\cQ\Psi +\frac{1}{3}\int 
\Psi\star\Psi\star\Psi\,\,,
\end{equation}
with the ``universal'' BRST operator $\cQ$ written in terms of
the ghost fields only. Under this assumption one may factorize
the string fields into the ghost and the matter parts,
$\Psi=\Psi^{matter}\otimes\Psi^{ghost}$
and the equation for the matter part becomes simplified,
\begin{equation}
 \Psi^{matter}\star \Psi^{matter} = \Psi^{matter}\,\,.
\end{equation}
This is the defining equation for the projection operators
with respect to Witten's star product $\star$.
Since the ghost part is universal, the solutions to
this equation are supposed to describe the various 
backgrounds of the open strings, namely D-branes.
Motivated by this observation there is a rapid development
in the construction of the projectors of the string field theory
\cite{r-RSZ3, r-GT1, r-KO, r-RSZ4, r-David, r-Matsuo1, 
r-GT2, r-RSZ5,r-Matsuo2}.

In particular there is direct argument \cite{r-RSZ4, r-Matsuo1}
in terms of the boundary conformal field theory that 
there are at least two types of the projection operators 
(the identity and the sliver states) associated with
every Cardy state which describes a D-brane. 
A critical property of Cardy state is that it has
a well-defined field content in the open string channel.
In particular, there is always an identity operator in the
channel connecting the same brane.
From the background independent formulation of the
open string field theory \cite{r-RZ}, one may always construct
the wedge state as follows,
\begin{equation}
 |n\rangle_a \equiv U_n|\mbox{vac};a,a\rangle\,\,,
\end{equation}
where $U_n$ describes the conformal transformations 
\cite{r-RZ, r-RSZ4},
\begin{equation}
 \tilde{z}=\tan\left(\frac{2}{n}\mbox{arctan}(z)\right),
\end{equation}
written in terms of the Virasoro generators.
$|\mbox{vac}; a, a\rangle$ is the vacuum state
in the open string sector with the both ends attached
to the same D-brane labeled as $a$. It was shown in \cite{r-RZ}
that they satisfy
\begin{equation}
 |n,a\rangle \star |m, a\rangle = |n+m-1,a\rangle\,\,.
\end{equation}
Clearly $n=1$ (the identity) and
$n=\infty$ (the sliver) give the projection operators.

The sliver state turns out to be
rank one and  considered as
a good candidate to describe the Chan-Paton factor of
a single D-brane.
On the other hand, in our previous work \cite{r-Matsuo2}, 
we indicated that the identity projector also enjoys
the following properties,
\begin{enumerate}
 \item It defines the projection to the open string states with
       both ends attached to a particular D-brane.  Namely if we
       denote $\cC$ as the algebra generated by the 
       open string fields with all possible boundary conditions,
       $\cI^a=|1\rangle_a$ picks up the open string field algebra
       $\cA^a$ with the particular boundary condition $a$.  Namely
\begin{equation}\label{e-identity_projector}
 \cI^a\star \cC \star\cI^a = \cA^a\,\,.
\end{equation}
 \item While the trace of $\cI^a$ is apparently infinite, there is
       a regularization which is common in BCFT that
\begin{equation}
 \mbox{Tr}\,(\,\cI^a) = \langle\mbox{vac} |a\rangle\,\,.
\end{equation}
Here $\langle \mbox{vac}|$ is the closed string vacuum
and $|a\rangle$ is the Cardy state.  The quantity 
in the right hand side is called as the boundary entropy
and can be identified with the D-brane tension \cite{r-HKMS}.
\end{enumerate}
Since these two properties are quite desirable, we
conjectured that the identity projector 
may be also a good candidate to 
describe the noncommutative soliton of the string 
field theory which describes D-branes.  

The use of the identity projector seems at least
puzzling from the viewpoint of
the standard conjecture \cite{r-RSZ2, r-RSZ3, r-GT1, r-RSZ4, r-GT2}.
Namely the rank of the projector is usually 
identified with the number of the D-branes.
This assumption stems from the fact that the potential energy 
\begin{equation}\label{e-potential}
 V=\frac{1}{2}\int \Psi^2 -\frac{1}{3}\int \Psi^3
\end{equation}
expanded around a particular projector $p$
has the $n\times n$ ($n$ is the rank of the projector $p$) 
negative modes and is identified as the number of
the tachyon states.  

Naive application of this idea to the identity projector
does not make any sense because the identity projector has
infinite rank.
However, we think that there is a possible
way out.  Namely in (\ref{e-potential}) one should add
the kinetic term.  Usually if the kinetic term is defined
in terms of $L_0$ (Virasoro generator), there is a contribution
to the mass term and the infinite degeneracy is resolved.

The problem is that in the pure ghost BRST operator there is no
matter Virasoro operator in the kinetic term. It means that we
can not use the above argument to save the identity projector.
However, we think that this is also a problem of the current
formulation of the
vacuum string field theory.  For example, in \cite{r-RSZ4},
it was observed that the expansion around D-brane background
(the sliver state) has no physical modes (every variation of the sliver 
can be absorbed by the gauge transformation).  In \cite{r-GT2}, it is also
indicated that the ghost action vanishes identically.
In any case, we think that there will be no hope to recover
the perturbative open string spectrum without introducing 
the matter Virasoro algebra at some point.

Since the appearance of the projector to describe D-branes
is quite natural from the viewpoint of K-theory 
\cite{r-Witten3,r-Matsuo0,r-HM},
we would like to try the other scenario, namely the
purely cubic action to describe the vacuum string theory.
By combining the idea of \cite{r-HLRS} with the identity
projector, we show that it gives a partially successful scenario
to this problem in a sense it reproduces the correct
perturbative spectrum on the survived D-branes.  While it
also gives at the same time some undesirable features, 
we think that it gives an intermediate  step to 
understand the correct description of the vacuum string field theory.

\section{SFT in the presence of several D-branes and split string method}

Before we discuss the vacuum string field theory, we would
like to give some definitions of the open string field theory
in the presence of several D-branes.  
Suppose we start from $N$ D-branes with the boundary condition
specified by the Cardy  state $|a\rangle$ $a=1,\cdots,N$\footnote{
One may take some of them to be identical D-brane. In that case
this label specifies Chan-Paton factor.}.
The open string field in this setting is described by
a $N\times N$ matrix $\Psi$ whose $ab$-th component is
written as $\Psi^{ab}$. $\Psi^{ab}$ describes the 
open string which connects D-branes  with labels $a$ and $b$.
We have to define the (perturbative or conventional)
BRST operator $Q$ for each sector by combining the Virasoro operator
of the matter sector (which is different 
for each sector) and the ghost sector which is universal.
We will use $Q_{ab}$ if we want to specify the operator
on a particular sector. We write the matrix $Q\Psi$ 
to mean the matrix whose $ab$-th component is
$Q_{ab}(\Psi^{ab})$.

Since the integration operator and the three string vertex 
can be written in the universal form \cite{r-RZ} 
(namely written in terms of Virasoro operators without 
specifying the particular  representation), the matrix
generalization of Witten's lagrangian is well-defined,
\begin{equation}
 S=\int \left(\frac{1}{2}\Psi \star(Q \Psi) + 
\frac{1}{3} \Psi\star\Psi\star\Psi
\right)\,\,,\quad
\int\equiv \sum_a \int_a
\end{equation}
Here the integration $\int_a$
is defined by the identity operator for the open string
in $aa$ channel. The summation with respect to 
$a$ gives the trace over Chan-Paton factor
when the D-branes are identical. 
The integration symbol $\int$ thus combines the trace of
matrix together with usual integration of the string field.
This action has the matrix generalization of the gauge symmetry,
\begin{equation}
 \delta \Psi^{ab}=Q_{ab} \epsilon_{ab} 
 + \sum_c (\epsilon_{ac}\star \Psi^{cb}-
\Psi^{ac}\star\epsilon_{cb})\,\,,
\end{equation}
where $\epsilon_{ab}$ is a string field from $ab$-sector.
Appearance of non-abelian gauge symmetry even for 
the intertwining strings might sound strange.  
However in these cases
$\epsilon_{ab}$ gives only the massive gauge
symmetry and does not appear in the massless sector.

In this language, the description of the identity projector
becomes very simple.  Namely for each $aa$-sector, one may define
the identity operator $|1\rangle_a\otimes \cI^{gh}$
($\cI^{gh}$ is the universal identity operator of the ghost field).
The projector $\cI^a$ is then
defined as putting it at $aa$-th component and putting zero
in other entries.  It is obvious that it satisfies 
\begin{equation}
 \cI^a\star\cI^b=\delta_{ab}\cI^b\,\,,\qquad
 \sum_{a}\cI^a=\cI_{\cC}
\end{equation}
($\cI_\cC$ is the identity for the whole system) 
and (\ref{e-identity_projector}).

For the discussion in the next section, it will be
useful to introduce the intuitive picture given by
the split string formalism \cite{r-GT1, r-GT2,
r-KO, r-AAB}.  Since the ghost sector
is common and the same for any boundary conditions we
concentrate our attention to the matter sector. We use
the notation of \cite{r-GT1, r-GT2} in the following.

We start from the discussion of $N$ D-25 branes
in the flat background.
Suppose that the open string field
$\Psi^{aa}$ whose both ends are connected to the $a$-th
($1\leq a \leq N$)
D-brane is described by the embedding function 
$x_{aa}^\mu(\sigma)$ ($0\leq \sigma \leq \pi$, $\mu=0,\dots,25$)
with the Neumann boundary condition at $\sigma=0,\pi$.
In the split string formalism, 
we divide it into the left and the right halves,
\begin{equation}
 \ell_a^\mu(\sigma)=x_{aa}^\mu(\sigma),\quad
 r_a^\mu(\sigma)=x_{aa}^\mu(\pi-\sigma),
 \quad (0\leq\sigma< \frac{\pi}{2}).
\end{equation}
The boundary condition at the midpoint should be
Dirichlet type in order to properly connect the left and the right halves.
$\ell(\sigma)$ and $r(\sigma)$ thus have the boundary condition
of Dirichlet-Neumann type and can be expanded as,
\begin{equation}
 \frac{\ell_a^\mu(\sigma)}{\sqrt{2}}=\sum_{n=0}^\infty \ell_{a, 2n+1}^\mu
\cos(2n+1)\sigma,\quad
 \frac{r_a^\mu(\sigma)}{\sqrt{2}}=\sum_{n=0}^\infty r_{a, 2n+1}^\mu
\cos(2n+1)\sigma.
\end{equation}
At the midpoint, since we fix $\ell(\frac{\pi}{2})=r(\frac{\pi}{2})=0$,
we need an extra degree of freedom (the zero mode) and we write it
as $\bar{x}$ (note that it does not have the index of the boundary).

To describe the open string that intertwine different D-branes
(say $a$ and $b$), we define the embedding function
by combining $\ell_a$ and $r_b$,
\begin{equation}\label{e-xab}
 x_{ab}^\mu(\sigma)=\left\{
\begin{array}{ll}
 \ell_a^\mu(\sigma)+\bar x\qquad\qquad&0\leq\sigma\leq \frac{\pi}{2}\\
 r_b^\mu(\pi-\sigma)+\bar x& \frac{\pi}{2}\leq \sigma \leq\pi
\end{array}
\right.
\end{equation}
Thanks to the Dirichlet boundary condition at $\sigma=\pi/2$,
such a mixed combination becomes consistent.
The string field for $a$-$b$ sector is defined as a
matrix indexed by $\ell_a$ and $r_b$
\begin{equation}
 \Psi^{ab}(x_{ab})\rightarrow \hat\Psi^{ab}(\ell_a,r_b)\,\,,
\end{equation}
and the multiplication is defined through the contraction,
\begin{equation}\label{e-multiplication}
 \hat\Psi^{ab}\star\hat\Phi^{bc}(\ell_a,r_b)
 = \int  \prod_{\sigma}dy(\sigma)
 \hat\Psi^{ab}(\ell_a,y_b)\hat\Phi^{bc}(y_b,r_c).
\end{equation}
The zero mode can be incorporated by multiplying
$e^{ip\bar x}$ if the momentum of the open string is $p$.

Up to this point, it is (at least intuitively) clear 
that our argument does not depend on the fact that we start from the 
$N$ identical D-25 branes.  Indeed the definition of
(\ref{e-xab},\ref{e-multiplication}) 
is always consistent even if we have the mixed boundary
conditions.
In this case, however, there is a subtlety in the definition
dynamical degree of freedom at the midpoint.  The dynamical
degree of freedom of the open string $x^{ab}$ belongs to the
twisted sector. We have to split it into two pieces and the question
is whether one may limit the boundary condition at $\sigma=\pi/2$ 
as Dirichlet.  In the generic BCFT, one may further need some
sort of discrete summation over sectors even at $\sigma=\pi/2$. 
Our argument in the next section will depend on the splitting
the dynamical degree of freedom to some extent.  To be strict,
it will be safer to assume that the discussion in the next section 
is made only for the $N$ identical D-branes while we will use
the universal notation.

In the split string formalism, the definition
of the identity projector becomes quite
intuitive.  We introduce the $N\times N$ 
matrix valued string field $p$. It is defined that we put
\begin{equation}
\cI_{gh}\otimes \pi_{ab}\prod_\sigma\delta(\ell_a(\sigma)-r_b(\sigma))
\end{equation}
at $a$-$b$ th entry where $\pi_{ab}$ is a constant complex number
and $\cI_{gh}$ is the identity operator of the ghost sector
\cite{r-AAB, r-GT2}.  By the matrix multiplication, it satisfies
$p^2=p$ if $\pi_{ab}$ satisfies $\sum_b\pi_{ab}\pi_{bc}=\pi_{ac}$.
We note that in the definition of the identity
operator, we have to equate the boundary conditions
in order that the delta function functional $\delta(\ell_a-r_b)$
is consistent.

We comment that in the tachyon condensation process from D-$(p+2)$ brane
to D-$p$ branes \cite{r-HKLM, r-Witten2} 
in the Seiberg-Witten limit \cite{r-SW}, we have already met
the matrix $\pi_{ab}$.  Indeed in this example one may take
$p+1$ directions to be Neumann for $\ell_a^\mu$ and $r_a\mu$
and the zero mode algebra gives the matrix $\pi_{ab}$.
It is the situation where the string field algebra
becomes the direct product of $\cal B(H)\otimes\cA$
where $\cA$ is the open string algebra between D-$p$ branes.
The matrix $\pi_{ab}$ can be precisely 
identified with GMS soliton \cite{r-Witten2} after
Weyl correspondence.  The appearance of GMS soliton
was also discussed in the direct string amplitude calculation
of $p-p'$ system in \cite{r-CIMM}.

\section{Identity projector in Purely Cubic Theory}

In the following, we consider the purely cubic system 
for the matrix extended open string fields,
\begin{equation}\label{e-cubic}
 S=\frac{1}{3}\int \Psi^3\,\,.
\end{equation}
Compared to (\ref{e-RSZ})
it corresponds to the choice 
$
 \cQ=0\,\,,
$
as the BRST operator which is 
in a sense universal (purely ghost).
This action was the first example of the background
independent formulation of the open string field theory
\cite{r-FWY,r-HLRS} 
(see also \cite{r-HIKKO, r-HMMW, r-Romans, r-Kluson}).

It was suggested \cite{r-RSZ1} that it will not likely
to be the closed string
vacuum which is discussed in the literature
\cite{r-Vacuum, r-Ohmori}.  Indeed we will meet a certain
difficulty which seems to be originated from 
this fact in our later discussion.

While these facts are discouraging,
it has certainly  a  merit that it was known already
\cite{r-HLRS,r-Romans} that one can 
reproduce the conventional perturbative string field theory
theory  as the fluctuation around a
solution to $\Psi^2=0$,
\begin{equation}\label{e-vacuum}
 \Psi_0= Q_L \cI\,\,,
\end{equation}
for the single component theory.
Here $\cI$ is the identity operator of the whole system
(matter+ghost) and $Q_L$ is the perturbative BRST current
integrated over the left half of the open string.
By putting $\Psi=\Psi_0+\psi$ into 
(\ref{e-cubic}) gives,
\begin{equation}
 S=\frac{1}{2} \int \psi \star Q \psi +\frac{1}{3} \int \psi^3\,\,,
\end{equation}
which is the correct open string field theory on a D-25 brane.

In the following, we would like to construct 
a similar solution in the multi-component theory.
Matrix generalization is essential to discuss
the tachyon condensation since in such theory we assume
that there are infinite number of D-branes which is annihilated
(or melted) in the closed string vacuum.  In a sense, the situation
is similar to consider ``Dirac sea'' for D-branes.
In order to discuss
the creation/annihilation of D-branes, we need to start from
the (in general infinite dimensional) matrix generalization.

Here we have a comment on the background independence of the string
field theory. In our formalism, we need to start from a particular
background for the closed string. From that data,  one can in principle
construct the possible boundary states and determine the matrix
algebra of the open strings.  Since we introduce the all possible
boundary conditions to construct $\Psi$, it is background independent
in the open string sense.  On the other hand, since we need the
information of the closed string background, it is not universal
in the closed string sense at least at the tree level.
We expect there are some consistency conditions for the
closed string background at the quantum level.

In this setup we argue that one may replace the identity
in (\ref{e-vacuum}) by the identity projectors defined 
in the previous section which is BRST invariant
\begin{equation}
 Q_{aa}(\cI^a)=0.
\end{equation}
Intuitively it describes a single D-brane
created from the vacuum.

In the following, we pick up one of the 
projector $\cI^a$ and write it as
$p$ and the sum of the all the rest as $q$. It is then 
trivial to show that 
\begin{equation}
 p\star p=p, \quad q\star q=q, \quad p\star q=0,
  \quad p+q=\cI_\cC\,\,.
\end{equation}
We start to prove that
\begin{equation}\label{e-QLp}
 \Psi_0 = Q_{aL} \, p\,\,,
\end{equation}
gives also a solution to $\Psi_0^2=0$.  
Here we assume that we can 
split the BRST charge for $ab$-sector as,
\begin{equation}
 Q_{ab}=Q_{aL}+Q_{bR}\,\,.
\end{equation}
Roughly speaking, $Q_{aL}$ (resp. $Q_{bR}$) is the BRST charge
for the half strings $\ell^a(\sigma)$ (resp. $r^b(\sigma)$).
We note however that they are not nilpotent.
As we have commented in the last section, this splitting
may be subtle for the open strings with twisted boundary conditions
because of the degree of freedom at the midpoint.
In any case, this splitting of BRST charge is essential
in our discussion in the following one needs modification
if there is a contribution from the midpoint.  For the
identical D-branes, we think that there will be no modification.

Under this caution, the calculation becomes
parallel to the original  \cite{r-HLRS} and
we  need (at least formally) the following identities
which was explicitly proved in \cite{r-Romans} for the
single component theory.
\begin{eqnarray}
 Q_{a,R} \,p &=& -Q_{a,L} \, p\label{e-a}\\
 (Q_{b,R} A_{ab})\star B_{bc} & = & - (-1)^{|A|} A_{ab}
\star Q_{b,L} B_{bc}\label{e-b}\\
 \left\{Q_{aa}, Q_{a,L}\right\} & = & 0\label{e-c}\,\,.
\end{eqnarray}


The proof of the nilpotency is given as follows.
First by using (\ref{e-a},\ref{e-b})
\begin{equation}
Q_L \, p \star Q_L \, p= -Q_R \, p\star Q_L\, p  = 
p\star Q_L^2 \, p.
\end{equation}
Second by using (\ref{e-b})
\begin{equation}
 Q_L \, p \star Q_L \, p= (Q_R Q_L\,  p)\star p.
\end{equation}
If we note that $Q_{L,R}$ does not change the boundary conditions,
$p\star Q_L^2 \ p= Q_L^2 \, p$ and 
$(Q_R Q_L\,  p)\star p=Q_R Q_L\,  p$.  Summing up these two formulae gives
\begin{equation}
  Q_L \, p \star Q_L \, p=\frac{1}{2} Q Q_L\,  p=
-\frac{1}{2} Q_L Q \, p=0.
\end{equation}
In this sense each identity projector indeed defines
a solution to the purely cubic action.

We have to note that,
unlike the situation in \cite{r-HLRS}, the derivative defined by
$Q_L p$ {\em does not} give BRST operator for the
string fields in the generic sectors.  
Indeed,
\begin{equation}
 D_{(Q_L \, p)} B = (Q_L \, p)\star B - (-1)^{|B|} B\star (Q_L \, p)
 = p\star(Q_L\, B)+ (Q_R\, B)\star p\,\,.
\end{equation}
If we restrict  the string field to $aa$-sector,
it gives of course the BRST operator. The existence of
other sectors breaks such a property. 
Accordingly the kinetic term of the action has
some undesirable terms.  If we put $\Psi=Q_L\, p+\psi$
into the action,
the contributions to the kinetic term takes the following form,
\begin{equation}\label{e-kinetic}
 \int \psi^2\star Q_L p = \frac{1}{2}\left(\int \psi_0 \star Q\psi_0
+\int t \star Q_L  \bar t + \int \bar t \star Q_R t
\right)\,\,.
\end{equation}
Here $\psi_0,t,\bar t$ are components of $\psi$,
\begin{equation}
 \psi_0=p\star\psi\star p,\quad
 t=q\star \psi\star p,\quad
 \bar t= p\star \psi\star q\,\,.
\end{equation}

The appearance of three terms in the kinetic terms can be
interpreted in the context of the tachyon condensation
\cite{r-Sen, r-HKLM, r-Witten2}  as follows.
We start from the D-brane system described by the
various boundary conditions where the intertwining open strings
describes the algebra $\cC$.  In this system, we consider the tachyon
condensation where only 
the D-branes picked up by the projector $p$ survives
after tachyon condensation.
The string field $\psi_0$ is then interpreted as the open strings
on the survived branes.  On the other hand,
two string fields $t,\bar t$ describe the intertwiner
between the survived and the disappeared ones.
Whereas we have a correct kinetic term  (\ref{e-kinetic})
for $a$, it is not reasonable that we have
a sort of (``incomplete'') kinetic terms for $t,\bar t$.

We would like to indicate that
this is actually a similar situation to
the field theory limit \cite{r-HKLM} for the noncommutative
tachyon.  Indeed, there also appeared the corresponding
components between the created and the melted D-branes.
These fields, however, are the analogue of W-bosons in the
system with spontaneously broken symmetry.
It was observed that they have the string scale
mass and infinitely degenerate. This puzzle was solved
\cite{r-GMS2,r-Seiberg, r-Sen2} by noting the correct choice
of variables and  vanishing of the overall factor of
the kinetic action.

In our case, one may also regard the intertwining modes
$t,\bar t$ as a kind of W-bosons for the symmetry breaking.
In the original action (\ref{e-cubic}), the gauge symmetry 
(gauge group on the configuration space of the open strings)
is generically $U(\infty)$ if we consider all possible
D-branes which acts on the string fields as,
\begin{equation}
 \delta_\epsilon \Psi=\epsilon\star\Psi-\Psi\star\epsilon\,\,.
\end{equation}
In the expansion around the classical solution (\ref{e-QLp}),
the gauge symmetry is transformed to,
\begin{equation}
 \delta\psi= (p\star (Q_{L,a}\epsilon)+(Q_{R,a}\epsilon)\star p)
+\epsilon\star \psi-\psi\star\epsilon\,\,.
\end{equation}
As we have commented, the first term can be written in
the form $Q\epsilon$ only when $\epsilon$ belongs to $aa$-sector.
If it belongs to the $oo'$-sector
with $o\neq a, o'\neq a$ then this term simply vanishes.
On the other hand, if it belongs to $ao$ or $oa$ with $o\neq a$
sectors, it can not be written as the conventional form of the
gauge transformation.  Therefore it may be adequate to consider that
the gauge symmetry $U(\infty)$ is broken to $U(1)\oplus U(\infty-1)$.

We may guess the off diagonal gauge transformation may
remove the unwanted degree of freedom. Indeed, when $t,\bar t$
is infinitely small, one may approximate the gauge transformation
with the explicit sector index as,
\begin{equation}
 \delta t_{oa}=Q_{R,a}\epsilon_{oa}\,\,,\quad
 \delta \bar t_{ao}= Q_{L,a}\epsilon_{ao}\,\,.
\end{equation}
While it is difficult to prove,
the use of this gauge transformaion may remove
the unwanted degree of freedoms.
We note that $Q_{L,R}^2$ is written  in terms of ghost 
fields only \cite{r-Romans}.

\section{Summary}
We summarize the results and the conjectures of this letter.
\begin{enumerate}
 \item From every identity projector associated with any D-brane,
one may construct the classical solution of 
the matrix version of the purely cubic open string field theory.
Since it contains all possible open string degree of freedom, 
one may call it as ``universal action''.
 \item The expansion around this classical solution gives a correct
action of the open strings on the created D-brane from the vacuum.
There exist some unwanted degree of freedom but it seems to be
possible to gauge them out.
\item As already noticed in \cite{r-RSZ1} and may be by others,
the purely cubic theory has an obvious drawback that the value
of the action $S(\Psi_0)$ for any classical solution $\Psi_0$
is always zero.  In our case, we are supposed to obtain the
D-brane tension from this factor but we can not.  
A possible resolution of this puzzle is that in the 
possible operation of
gauging away $t,\bar t$, we will have a singular transformation
that will reproduce a nontrivial contribution to the action.
This is, of course, rather hard to confirm.  There will be
probably a framework where this effect is automatically produced
as Rastelli, Sen and Zwiebach endeavored.
\item In any case, our work may suggest that one needs 
a singular classical solution to have a nontrivial 
BRST cohomology. We think the magic, if any,
can only come from the serious examination of
the notorious midpoint degree of freedom.
\end{enumerate}

\vskip 5mm
\noindent{\em Acknowledgement:} 

The author would like to thank T. Kawano, and K. Ohmori for
discussions and K. Fujikawa for encouragements.

The author is supported in part by Grant-in-Aid (\#13640267)
and in part by Grant-in-Aid for Scientific Research
in a Priority Area ``Supersymmetry and Unified Theory of 
Elementary Particle'' (\#707) from the Ministry of Education,
Science, Sports and Culture.

%


\begin{thebibliography}{99}
\bibitem{r-RSZ1} L. Rastelli, A. Sen and B. Zwiebach,
``String Field Theory Around The Tachyon Vacuum'',
{\tt hep-th/001225}.
\bibitem{r-RSZ2} L. Rastelli, A. Sen and B. Zwiebach,
``Classical Solutions in String Field Theory Around
the Tachyon Vacuum'', {\tt hep-th/0102112}.
\bibitem{r-Witten} E. Witten, 
Nucl. Phys. B268 (1986) 253.
\bibitem{r-RSZ3} L. Rastelli, A. Sen and B. Zwiebach,
``Half-strings, Projectors and Multiple D-branes in Vacuum String
Field Theory'', {\tt hep-th/0105058}.
\bibitem{r-GT1} D. J. Gross, W. Taylor,
``Split string field theory I'', {\tt hep-th/0105059}.
\bibitem{r-KO} T.~Kawano and K.~Okuyama,
``Open string fields as matrices,''
hep-th/0105129.
\bibitem{r-RSZ4} L. Rastelli, A. Sen, B. Zwiebach,
``Boundary CFT Construction of D-brane in Vacuum String
Field Theory'', hep-th/0105168.
\bibitem{r-David} J. R. David 
``Excitations on wedge states and on the sliver'',
hep-th/0105184.
\bibitem{r-Matsuo1} Y. Matsuo, ``BCFT and Sliver state'',
{\tt hep-th/0105175}.
\bibitem{r-GT2} D. J. Gross, W. Taylor,
``Split string field theory II'', {\tt hep-th/0106036}.
\bibitem{r-RSZ5} L. Rastelli, A. Sen, B. Zwiebach,
``Vacuum string field theory,''
{\tt hep-th/0106010}.
\bibitem{r-Matsuo2}
Y.~Matsuo, ``Identity projector and D-brane in string field theory,''
{\tt hep-th/0106027}.
\bibitem{r-RZ} L. Rastelli and B. Zwiebach, ``Tachyon
potentials, star products and universality'', {\tt hep-th/0006240}.
\bibitem{r-HKMS} J. Harvey, S. Kachru, G. Moore and E. Silverstein,
JHEP 0003 (2000) 001, {\tt  hep-th/9909072}.
\bibitem{r-Witten3} E. Witten, 
JHEP {\bf 9812} (1998) 019, {\tt hep-th/9810188}.
\bibitem{r-Matsuo0} Y. Matsuo, 
Phys.\ Lett.\ B {\bf 499} (2001) 223
, {\tt hep-th/0009002}.
\bibitem{r-HM} J. A. Harvey and G. Moore,
``Noncommutative tachyons and K-theory,''
{\tt hep-th/0009030}.
\bibitem{r-HLRS} G.~T.~Horowitz, J.~Lykken, R.~Rohm and A.~Strominger,
Phys.\ Rev.\ Lett.\  {\bf 57} (1986) 283.
\bibitem{r-AAB} A. Abdurrahman, F. Anton and J. Bordes,
Nucl. Phys. B397 (1993) 260--282, Nucl. Phys. B411 (1994) 693--714.
\bibitem{r-HKLM} J. A. Harvey, P. Kraus, F. Larsen and E. J. Martinec,
JHEP {\bf 0007} (2000) 042, {\tt hep-th/0005031}.
\bibitem{r-Witten2} E.~Witten,
``Nonocommutative Tachyons and String Field Theory''
{\tt hep-th/0006071};\\
E. Witten, Int. J. Mod. Phys. A16 (2001) 693-706,
{\tt hep-th/0007175}.
\bibitem{r-SW} N. Seiberg and E. Witten,
JHEP {\bf 9909} (1999) 032, {\tt hep-th/9908142}.
\bibitem{r-CIMM} 
B.~Chen, H.~Itoyama, T.~Matsuo and K.~Murakami,
Prog.\ Theor.\ Phys.\  {\bf 105} (2001) 853
{\tt hep-th/0010066};\\
K.~Murakami,
``p-p' system with B field and projection operator noncommutative  solitons,''
{\tt hep-th/0104243}.
\bibitem{r-FWY} Unpublished and independent
observations by D. Friedan, E. Witten,  T. Yoneya (1985)
and probably by others.
\bibitem{r-HIKKO} H.~Hata, K.~Itoh, T.~Kugo, H.~Kunitomo and K.~Ogawa,
Phys.\ Lett.\ B {\bf 175} (1986) 138.
\bibitem{r-HMMW} G. T. Horowitz, J. Morrow-Jones, S. P. Martin
and R. P. Woodard, Phys. Rev. Lett. 60 (1988) 261.
\bibitem{r-Romans} L. J. Romans, 
Nucl. Phys. B298 (1988) 369-413.
\bibitem{r-Kluson} J. Kluson, ``Proposal for Background Independent
Berkovits' Superstring Field Theory'', {\tt hep-th/0106107}.
\bibitem{r-GMS2} R. Gopakumar, S. Minwalla and A. Strominger,
JHEP 0104 (2001) 018, {\tt hep-th/0007226}.
\bibitem{r-Seiberg} N. Seiberg, JHEP 0009 (2000) 003, 
{\tt hep-th/0008013}.
\bibitem{r-Sen2} A. Sen, JHEP 0011 (2000) 035, {\tt hep-th/0009038}.
\bibitem{r-Vacuum} Some of the pioneering references are:\\
V. A. Kostelecky and S. Samuel, Phys. Lett. B207 (1988)  169;\\
A. Sen, JHEP 9912 (1999) 027, {\tt hep-th/9911116};\\
A. Sen, B. Zwiebach, JHEP 0003 (2000) 002, {\tt hep-th/9912249}.
\bibitem{r-Ohmori} K. Ohmori, ``A Review on Tachyon Condensation
in Open String Field Theory'', {\tt hep-th/0102085}.
%
\bibitem{r-Sen} A. Sen, ``Non-BPS states and branes in string theory'',
{\tt hep-th/9904207} and references therein.
\end{thebibliography}
\end{document}